\documentclass[aps,pre,reprint,superscriptaddress,longbibliography,showpacs]{revtex4-2} %aps,pre,reprint,superscriptaddress,longbibliography,showpacs
\usepackage{amsmath,amssymb,graphicx,ragged2e }
\usepackage[table]{xcolor}
\newcommand\jl[1]{ {\color{blue} #1} }
\usepackage{multirow}
\usepackage{subcaption}
\usepackage{upgreek}
\usepackage[font=small,labelfont=bf, justification=centerlast, format=plain,belowskip=0pt]{caption}

\usepackage{times,bm,url}
\usepackage[colorlinks=true,breaklinks=true,allcolors=blue]{hyperref}
\newcommand{\orcidicon}[1]{\href{https://orcid.org/#1}{\includegraphics[height=\fontcharht\font`\B]{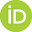}}}

\newcommand{\ex}[1]{\ensuremath{\langle #1 \rangle}} %expectation value

\DeclareMathOperator{\tr}{{Tr}}
\DeclareMathOperator{\sgn}{{sgn}}

\usepackage[scr=boondox]{mathalfa}
\def\ggg{\ensuremath{\mathscr{g}}}

\def\www{\ensuremath{{\mathscr{ w}}}}
\def\Oo{\ensuremath{{\cal O}}} %Order sign
\def\Jj{\ensuremath{{\cal J}}} %effective coupling

\def\AdS{\ensuremath{\text{AdS}}}  
\def\Ss{\ensuremath{{\cal S}}} 
\def\Hh{\ensuremath{{\cal H}}} %the Hamiltonian
\def\Nn{\ensuremath{{N}}}

\def\Ii{\ensuremath{{\cal I}}}

\def\Ggg{\ensuremath{\mathscr{G}}}
\usepackage[scr=boondox]{mathalfa}
\def\ggg{\ensuremath{\mathscr{g}}}

\def\hh{\ensuremath{\mathscr{h}}}

\def\Tt{\ensuremath{{\cal T}}} %iets
\def\Gg{\ensuremath{{\cal G}}} %nog iets
\def\Qq{\ensuremath{{\cal Q}}}

\def\p{\ensuremath{{\partial}}}
\def\i{\ensuremath{{\imath}}}

\newenvironment{claim}[1]{\par\noindent\underline{Claim:}\space#1}{}
\usepackage{xparse}
\DeclareDocumentCommand{\nint}{ O{} O{} m }{\ensuremath{ \int_{\mbox{\scriptsize $#1$}}^{\mbox{\scriptsize$#2$}}\!\!\! \mbox{\small $\,\mathrm{d}#3$\! }}}
\usepackage{microtype}

\begin{document}
	\title{Shared universality of charged black holes and the complex large-$q$ Sachdev-Ye-Kitaev model}
	
	\author{Jan C.\ Louw \!\orcidicon{0000-0002-5111-840X}}
	\affiliation{Institute for Theoretical Physics, Georg-August-Universit{\"a}t G{\"o}ttingen,  Friedrich-Hund-Platz 1, 37077~G{\"o}ttingen, Germany}
	
	\author{Stefan Kehrein}
	\affiliation{Institute for Theoretical Physics, Georg-August-Universit{\"a}t G{\"o}ttingen,  Friedrich-Hund-Platz 1, 37077~G{\"o}ttingen, Germany}
	
	\date{\today}
	\begin{abstract}
		We investigate the charged $q/2$-body interacting Sachdev-Ye-Kitaev (SYK) model in the grand-canonical ensemble. By treating $q$ as a large parameter, we are able to analytically study its phase diagram. By varying the chemical potential or temperature, we find that the system undergoes a phase transition between low and high entropies, in the maximally chaotic regime. A similar transition in entropy is seen in charged AdS black holes transitioning between a large and small event horizon. Approaching zero temperature, we find a first-order chaotic-to-non-chaotic quantum phase transition, where the finite extensive entropy drops to zero. This again has a gravitational analogue---the Hawking-Page (HP) transition between a large black hole and thermal radiation.
		
		An analytical study of the critical phenomena associated with the continuous phase transition provides us with mean field van der Waals critical and effective exponents. We find that all analogous power laws are shared with several charged AdS black hole phase transitions. Together, these findings indicate a connection between the charged $q\to\infty$ SYK model and black holes. 
	\end{abstract}
	
	\maketitle

	The analysis of condensed matter systems lacking quasi-particles is hindered by the unamenability of Fermi-liquid theory. One successful approach is via a class of disordered  Sachdev-Ye-Kitaev (SYK) models \cite{Sachdev2015,Fu2018}, or their related disorder-less planar/tensor matrix models \cite{Azeyanagi2018Feb,Witten2016Oct}.  Despite their non-integrability, one may find exact relations between the self-energy and Green's function $\Ggg$ \cite{Sachdev1993May}. This reduces the exponential complexity of the problem to a single Dyson equation purely in terms of $\Ggg$.  Although some analytical results exist in the infrared limit \cite{Maldacena2016Nov}, the full solutions are obtained numerically. 
	
	There is also a framework in which one may find exact analytical solutions. This is by considering $q/2$-body interactions, for large $q$, and treating $1/q$ as an expansion parameter. In this work, we present a study of such a model
	\cite{Fu2018,Louw2022Feb}
	\begin{equation}
	\Hh = J \sum_{\substack{1\le i_1< \cdots < i_{q/2}\le \Nn \\ 1\le j_1<\cdots< j_{q/2}\le\Nn}} X^{ i_{1}\cdots  i_{q/2}}_{j_{1}\cdots j_{q/2}} c^{\dag}_{ i_1} \cdots c^{\dag}_{ i_{\frac{q}{2}}} c_{j_{\frac{q}{2}}}^{\vphantom{\dag}} \cdots c_{j_1}^{\vphantom{\dag}}, \label{H}
	\end{equation}
	with a conserved U$(1)$ charge density $\hat{\Qq} = \frac{1}{N}\sum_i c_i^\dag c_i - 1/2$, with expectation values $\Qq \in [-1/2,1/2]$. Here $c^\dag, c$ are Dirac/complex fermionic creation and annihilation operators, respectively.  We will study this model in the grand-canonical ensemble with partition function $Z = \tr\{e^{- \beta [\Hh -\mu \Nn\hat{\Qq}]}\}$. The couplings, $X$, are complex random variables with zero mean, and a variance $\overline{\,|X|^2} =  [q^{-1} (q/2) !]^2 [2/\Nn]^{q-1}$. Such models have the advantage of being amenable to analytical solutions.  At neutral charge, $\Qq=0$, its thermodynamics reduces to its Majorana ($c^\dag = c$) counterpart  \cite{Maldacena2016Nov}. The inclusion of non-zero charge brings \eqref{H} in closer contact with electronic systems \cite{Sachdev2015,Zanoci2022Apr,Song2017}. By varying a chemical potential $\mu$, the conjugate to $\Qq$, we find that this model exhibits a phase transition similar to its finite $q$ equivalents \cite{Azeyanagi2018Feb,Ferrari2019Jul}. In contrast to the numerical results in the finite $q$ case, we are able to analytically study its phase diagram in the large $q$ limit. This is done by considering suitable polynomial scaling (in $q$) thermodynamic variables such as $T,\mu$. Such $q$-scalings have also previously been considered for two coupled Majorana SYK models \cite{Maldacena2018Apr} in the $q\to\infty$ limit. The analytical solutions to the equilibrium Green's functions $\Ggg$ and a proof that they remain valid for our considered scaling is given in App. \eqref{cl1}. 
	
	The Green's functions are key to studying the phase diagram. This is because the Kubo-Martin-Schwinger (KMS) relation, $\Ggg(\tau+\beta) = -e^{-\beta \mu}\Ggg(\tau)$ \cite[App. B]{Sorokhaibam2020Jul}, allows one to extract the exact equation of state \cite[eq.(43)]{Louw2022Feb}
	\begin{equation}
	\mu(\Qq) = 2 T \tanh^{-1}(2\Qq)+4\Qq \Jj(\Qq) \sin(\pi v/2)/q, \label{ChemPot}
	\end{equation}
	for large $q$, with effective coupling strength
	\begin{equation}
	\Jj(\Qq) \equiv J [1-4\Qq^2]^{(q-2)/4} \label{efCoup}
	\end{equation}
	and Lyapunov exponent $\lambda_L = 2\pi v T$, found by solving $\Jj/T = \pi v \sec(\pi v/2) $ \cite{Maldacena2016Nov}.  Note that, for any non-zero $\Qq = \Oo(q^0)$, the interaction is suppressed for large $q$, $\Jj \xrightarrow{q\to \infty} 0$. 
	
	One is able to retain non-zero (constant) effective coupling $\Jj$, by adjusting the model to have $\Qq$ dependent coupling $J(\Qq) \propto [1-4\Qq^2]^{(2-q)/4}$ \cite{Davison2017}. This allows the interactions to remain relevant at all charge densities. Notice, however, that with this adjustment, the Hamiltonian inherits a temperature dependence from the charge density \cite{Zanoci2022Apr}. In this case, there is no phase transition. In contrast, by a suitable $q$-dependent rescaling introduced below, we will find that the phase transition of the Hamiltonian \eqref{H}, without adjustments, persists even for $q\to\infty$.
	
	In particular, we find a van der Waals (vdW)-like phase diagram \cite{Cho2018,Kubiznak2012Jul}, with a line of first-order phase transition terminating at a critical end-point, where the transition is continuous. Associated with this are multiple power laws, the critical exponents of which we are able to calculate analytically. 
	Comparing exponents, we find that our model shares such a universality class with a wide range of models, including numerous AdS black holes, a non-exhaustive list of which is	\cite{Kubiznak2012Jul,Kubiznak2012Jul,Majhi2017Oct,Dolan2016May,Cao2021Mar,Dehyadegari2019Apr,Mandal2016Sep,Chamblin1999Aug,Niu2012Jan}. These similarities between black holes and the complex large-$q$ SYK model extends even beyond the shared universality class. For instance, over the phase transition, there is a drop in entropy reminiscent of the large-to-small horizon transition in Reissner-Nordstr{\"o}m (RN), charged and non-rotating, black holes. Such systems also appear in the study of non-Fermi-liquids, under the name  RN metals \cite{Zaanen2015Nov}.

\begin{figure}[h]
		\includegraphics[width=1\linewidth]{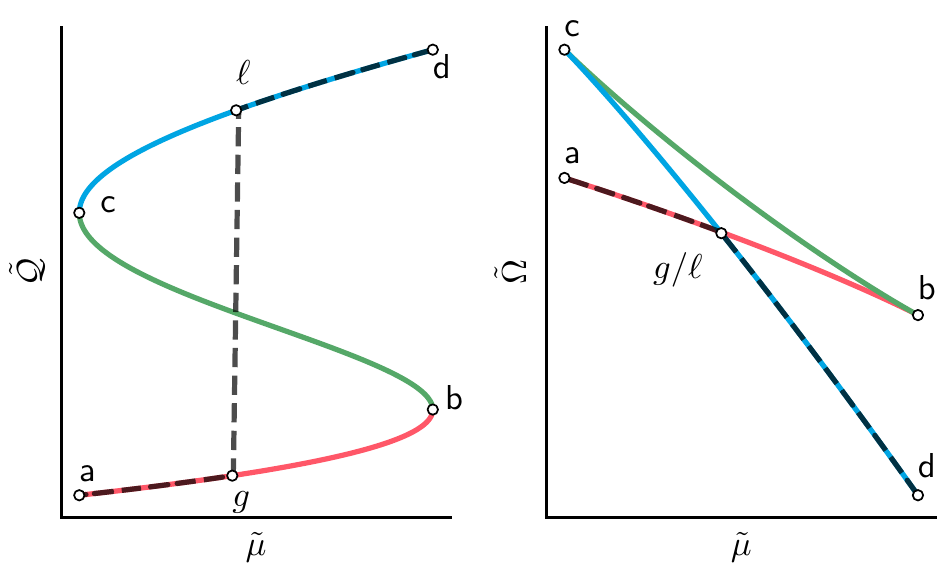}
		\caption{The three solutions to \eqref{EOS2} distinguished by color, for $\tilde{T}<\tilde{T}_\text{crit}$. The dashed line indicates the thermodynamically favorable solution. \hspace*{\fill}}
		\label{fig:chempot}
	\end{figure}

	\emph{Phase diagram.---} We start our analysis by considering two extremes. At zero charge density we are left with a strongly interacting pure Majorana-like SYK model,	while at any finite charge density $\Qq = \Oo(q^0)$ the interaction \eqref{efCoup} is trivial $\Jj\to 0$ leaving a Fermi gas. Somewhere in between these two extremes must lie a regime where interactions and density terms in \eqref{ChemPot} compete in a non-trivial way. Indeed, such a competition is found for thermodynamic quantities which scale like
	\begin{equation}
	T = \tilde{T} q^{-1}, \quad \mu = \tilde{\mu}q^{-3/2}, \label{scaling}
	\end{equation} 
	where tilde'd quantities are held fixed as $q\to\infty$. In turn, the charge densities scales like $\Qq = \tilde{\Qq} q^{-1/2}$, hence yielding a finite effective interaction \eqref{efCoup}, $\Jj(\Qq) \xrightarrow{q\to\infty} e^{-\tilde{\Qq}^2} J$. This scaling corresponds to the maximally chaotic regime, where $v = 1 - 2 \tilde{T}/(q \Jj) + \Oo(1/q^2)$ saturates the universal (chaos) bound $\lambda_L\to 2\pi T$ \cite{Maldacena2016}. Using the scaling \eqref{scaling} thus simplifies equation of state \eqref{ChemPot} to 
	\begin{equation}
	\tilde{\mu}(\tilde{\Qq}) = 4 \tilde{\Qq} [\tilde{T} + J e^{-\tilde{\Qq}^2}]+\Oo(1/q). \label{EOS2}
	\end{equation}
	Plotting this, as in fig. \ref{fig:chempot}, one notes that there exists a critical temperature, $\tilde{T}_{\text{crit}}= 2 J e^{-3/2}$, below which three solutions exist instead of one. The solution from b-c has unphysical negative compressibility. Such behavior is also seen in the vdW liquid-gas transition. Due to the difference in (charge) density, and similarities to the vdW system, we shall refer to the two physical solutions as the gaseous and liquid phases, given in pink and blue respectively.
	
	Out of the three, the thermodynamically favorable solution corresponds to the smallest grand potential (per lattice site) $\Omega = -T\ln Z/\Nn$. Considering the partition function, we observe the following relations $J \p_{J} \Omega = E$ and $\p_\mu \Omega = -\Qq$, where the energy density is given by \cite{Louw2022Feb} $$E =  -2(1-4\Qq^2)\Jj \sin(\pi v/2)/q^2,$$ which defines a set of differential equations. From the relation $\Omega = E - \mu \Qq - T \Ss$,  we observe that the entropy density $\Ss$ is the only unknown. We may find $\Ss$ by solving the set of differential equations for the scaling \eqref{scaling}, i.e., in the maximal chaotic regime $v\to 1$, $\p_{\tilde{\mu}} q^2 \Omega = -\tilde{\Qq}$ and $\p_{J} q^2 \Omega = -2e^{-\tilde{\Qq}^2}$. One may then verify by substitution that $\Ss = \ln 2-2 \tilde{\Qq}^2/q + \Oo(1/q^2)$, where the constant is found by using the free fermion solution at $J=0$. The corresponding grand potential is the written as $q^2 \Omega = \tilde{\Omega} -q \tilde{T}\ln 2$, with %even simpler is to use the thermodynamic identitiy \cite{Sachdev2015} $(\p_\Qq \Ss)_T = - (\p_T \mu)_\Qq$
	\begin{equation}
	\tilde{\Omega} \equiv -2 \tilde{\Qq}^2 \tilde{T} -2[1+2 \tilde{\Qq}^2]  J e^{-\tilde{\Qq}^2} +\Oo(1/q). \label{grandPot}
	\end{equation}
	
\begin{figure}
		\centering %, Metastable region $\Omega^{(2)}(\Qq) \ge 0$,
		\includegraphics[width=0.9\linewidth]{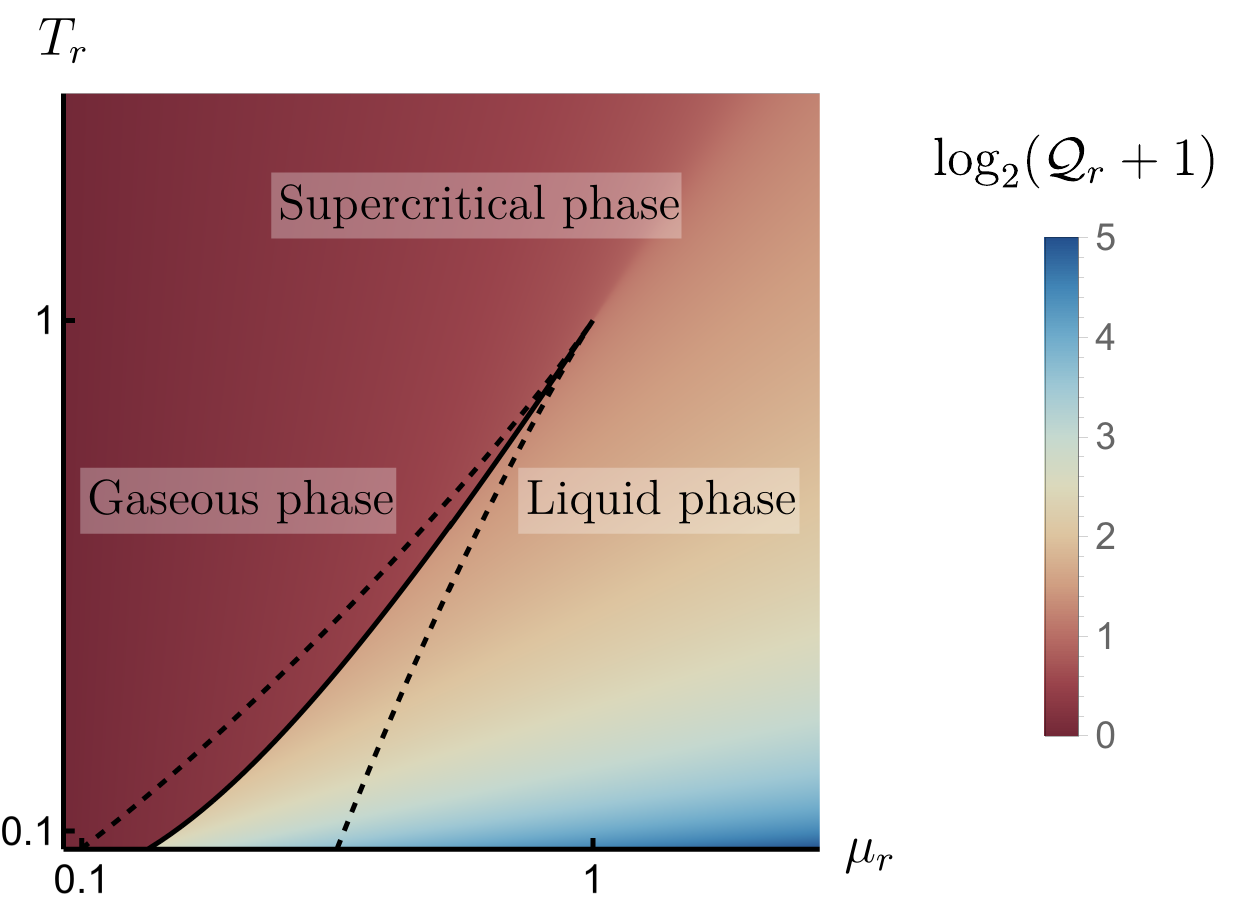} %in the $(\mu_r,T_r)$ plane
		\caption{Phase diagram, for scaling \eqref{scaling}, in terms of reduced variables, $T_r = \tilde{T}/\tilde{T}_{\text{crit}}$, $\mu_r = \tilde{\mu}/\tilde{\mu}_{\text{crit}}$ and $\Qq_r = \tilde{\Qq}/\tilde{\Qq}_{\text{crit}}$. The solid and dashed lines denote the coexistence and stability curves, respectively. The color corresponds to the thermodynamically favorable charge density.  \hspace*{\fill}}
		\label{fig:qmut} %Note how similar this diagram looks to that of \cite{Banados2017Aug}
	\end{figure}

	Considering $\tilde{\Omega}$ for these three solutions, plotted in fig. \ref{fig:chempot}, we find that the favorable charge density necessarily jumps between $g$ and $\ell$, missing the unphysical solution.  Also note the distinct swallowtail shape, from catastrophe theory, which is also common to RN phase transitions. This shape is indicative of a first-order phase transition, in this case from a low (charge) density gaseous phase to the dense liquid phase. This is induced either by increasing the chemical potential or decreasing the temperature, as seen from the phase diagram in fig. \ref{fig:qmut}. The stability limit curves, enclosing the region where both phases can coexist, coincide at the critical point $(\tilde{\mu}_{\text{crit}},\tilde{T}_{\text{crit}})$, with charge density $\tilde{\Qq}_{\text{crit}} = \sqrt{3/2}$. Here the two turning points b and c merge into an inflection point, where there is a continuous (second-order) phase transition. Above this lies the supercritical phase, identified by a single unique solution.

	%coexistence = binodal, stability limit = spinodal
	%Both turning points b,c are fixed points, i.e., points where $\Omega^{(1)}(\Qq)$ are zero. The first is a stable fixed point, however, the global minimum is to the right. The second is an unstable fixed point.
	
	\emph{Universality.---} Approaching the critical point, i.e., for small shifted variables $m \equiv \mu_r -1$, $\rho \equiv \Qq_r-1$, and $t \equiv T_r -1$, we find that various thermodynamic quantities display power laws. To study this, we shift and rescale the grand potential \eqref{grandPot} and consider it close to this point
	\begin{equation} %\frac{\tilde{\Omega} }{\tilde{\mu}_\text{crit}\tilde{\Qq}_\text{crit}} \bigg\vert^{\tilde{\Qq}=\tilde{\Qq}}_{\tilde{\Qq}=\tilde{\Qq}_\text{crit}} +m
	f = \frac{\tilde{\Omega} - \tilde{\Omega}_{\text{crit}} }{\tilde{\mu}_\text{crit}\tilde{\Qq}_\text{crit}}+t/3 +m= -\rho^2 \frac{t +(3\rho/2)^2}{3}+\Oo(\rho^5) \label{resf}
	\end{equation}
	which satisfies $\p_m f = -\rho$, stemming from $\p_{\tilde{\mu}} \tilde{\Omega} = - \tilde{\Qq}$. Further, the linear shift $t$ will not affect the specific heat.
	
	In particular, we focus on the minimized grand potential corresponding to the dashed line in fig. \ref{fig:chempot}. One may show that in terms of some ordering field $\hh$, it is homogeneous
	$f(t,\hh) = t^{2-\alpha} f(1,\hh t^{\beta -2+\alpha})$, where $\alpha$ and $\upbeta$ are the critical exponents characterizing the power laws. Models which share the same scale-invariant form $f$ under the renormalization group flow are said to belong to the same universality class. 
	
	The homogeneity property is satisfied for the ordering field $\hh = m - 2 t/3$, which restricts the form of $\hh$ up to a scaling constant. This leaves the  singular function
	$$f(t,\hh) = - \frac{|t|^{2-\alpha}}{6} - |\hh| |2t/3|^{\upbeta} -  \frac{3\hh^2}{2} |t|^{-\gamma}  + \Oo(\hh^3 |t|^{-5/2})$$
	for small $\hh$ and  $f(t,\hh) =
	- \frac{3}{4} |\hh|^{1 + 1/\delta}(1 +\Oo(t \hh^{-2/3}))$ for small $t$. The details of this calculation are given in the App. \ref{cl3}.  Here $\alpha$, $\upbeta$, $\gamma$, $\delta$ are the classical mean field critical exponents, given in table \ref{meanfield}. As such, our model falls into the vdW universality class. From $\rho(t,\hh) =-\p_{\hh} f(t,\hh)$, we have $\rho(0,\hh) =\hh^{1/\delta}$ and 
	\begin{equation}
	\rho(t,\hh) =\sgn(\hh)|2t/3|^{1/2} +  2\hh |2t/3|^{-1}  + \Oo(\hh^2 t^{-5/2})
	\label{rhosim}
	\end{equation}
	for small $\hh$.	
	The remaining critical exponents characterize the power laws along the line $\hh=0$: the order parameter $\rho(t,0) \propto |t|^{\upbeta}$, specific heat $C_\hh \propto -\p_{t}^2 f(t,0) \propto |t|^{-\alpha}$ and susceptibility $\chi_\hh \propto \p_{\hh}^2 f(t,\hh) \mid_{\hh=0}\propto |t|^{-\gamma}$.
	
	\begin{table}[h!]
		\caption{Table of critical exponents \label{meanfield}}
		\begin{tabular}{|c c c c |} 
			\hline
			$\alpha$  & $\upbeta$ & $\gamma$ &  $\delta$\\
			\hline
			$0$ & $1/2$ & $1$  & $3$\\ 
			\hline
		\end{tabular}
	\end{table}

	\emph{Effective exponents.---} The particular power law can be dependent on the line along which the critical point is reached. This feature is due to the mixing of chemical potential and temperature in the ordering field $\hh$ \cite{Wang2007May}. Since the corresponding exponents do not enter into the scale-invariant form of the model, they are not the critical exponents which define the universality class. However, these \emph{effective} exponents still describe physically relevant processes. As an example, let us consider the specific heat. For constant $\mu$, we have $C_{\mu} \propto -\p_t^2 f(t,-2t/3) \propto |t|^{-2/3}$, i.e. $\alpha_\mu = 2/3$. Here, the subscripts indicate which quantity is set to its critical value.	In contrast, for constant $\Qq$, we use the identity $C_\Qq = (\p_{T} E)_\Qq$. In the chaotic regime, the energy behaves as $E \propto T^2+\text{const.}$, leaving $C_\Qq \propto T \sim t^0$. While often associated with Fermi-liquid behavior, such a Sommerfeld, linear in $T$, specific heat also appears in RN \cite{Zaanen2015Nov}  and cuprate \cite{Loram1993Sep,Michon2019Mar,Legros2019Feb} strange metals for a range of doping levels. 
	The remaining \emph{effective} exponents can be obtained from \eqref{rhosim}, and are listed in tables II.\hyperref[a]{(a)} and II.\hyperref[b]{(b)}.

	\begin{table}[h]
		\caption{Tables of effective exponents   \hspace*{\fill}
		}
		\begin{minipage}[c]{0.02\columnwidth}
			\caption*{
				(a) %fixed field
			} \label{a} 
		\end{minipage}
		\begin{minipage}[c]{0.44\columnwidth}
			\begin{tabular}{|c c c |} 
				\hline
				$\alpha_\mu$ & $\upbeta_\mu$ & $\gamma_\mu$ \\
				\hline
				$2/3$ & $1/3$ & $2/3$ \\ 
				\hline
			\end{tabular}
		\end{minipage}\hfill
		\begin{minipage}[c]{0.02\columnwidth}
			\caption*{
				(b) %fixed order parameter
			} \label{b}
		\end{minipage} 
		\begin{minipage}[c]{0.44\columnwidth}
			\begin{tabular}{|c  c |} 
				\hline
				$\alpha_\Qq$   & $\gamma_\Qq$ \\
				\hline
				$0$  & $1$\\ 
				\hline
			\end{tabular}
		\end{minipage}
		%\vspace{-8pt}
	\end{table}
	
	%\emph{Universality.---} While the full grand potential is analytic, the minimized grand potential $\tilde{\Omega}^*$ is singular at the transition point \cite{Kardar2007}. In other words, it is non-differentiable over the coexistence line where $\rho_\ell \to \rho_g$. As such the singular behavior resides in the $\rho$ dependent parts. We consider $\tilde{\Omega}^*$ as a function $t$ and some ordering field $\hh$, which is a function of $t,m$. As such the non-singular part is a function of $t,\hh$, which we subtract off of $\tilde{\Omega}^*$. 

	\emph{Relation to gravity.---} By comparing order parameters and their conjugates, one can make various analogies between the models listed in table \ref{tab:title}. The similarities are strongest when comparing our model to RN black holes. These systems are defined by a charge $q_B$, an event horizon radius $r$ and electrical potential $\Phi = q_B r^{2-d}$ \cite{Niu2012Jan}. One may also consider such systems in an \emph{extended} $\AdS_{d+1}$ space \cite{Kubiznak2012Jul,Kubiznak2017Feb}, where the cosmological constant $\Lambda$ and its conjugate quantity, the volume $V$, are treated as thermodynamic variables. Here $V$, like in the vdW case, is the order parameter, while $-\Lambda$ acts as the pressure term $P$. 
	
	\begin{table}[h]
		\centering
		\captionof{table}{Analogies between models with shared universality class. \hspace*{\fill}} \label{tab:title} 
		\begin{tabular}{|c|c | c  |c| c |c |} 
			\hline
			Model& SYK & vdW  & \multicolumn{2}{c|}{RN-AdS$_{}$} \\
			\hline
			&	\cite{Louw2022Feb} & \cite{Cho2018,Kubiznak2012Jul} &  \cite{Kubiznak2012Jul,Kubiznak2017Feb} & \cite{Dolan2016May}  \\ 
			\hline
			order parameter &	$\Qq$ & $V$  & $V$ & $\Phi$ \\
			conjugate &	$\mu$ & $P$ &  $-\Lambda$ & $q_B$\\
			\hline
		\end{tabular}
	\end{table}

	These analogies are quantitative in the sense that all effective exponents also match. By this, we mean that by keeping order parameters fixed,  both exponents match II.\hyperref[b]{(b)}. Then, while keeping the conjugates fixed, we find three exponents matching II.\hyperref[a]{(a)}.  All models listen in table \ref{tab:title} also share the same critical exponents. As such, our model, the vdW liquid \cite{Kubiznak2012Jul} and multiple RN $\AdS_{d+1}$ black holes \cite{Kubiznak2012Jul,Kubiznak2017Feb,Dolan2016May,Niu2012Jan} all share a universality class, as well as having the same effective exponents.
	%Majhi2017Oct
	
	Besides sharing a universality class, these analogous models also have an abundance of qualitative commonalities. This is particularly apparent at low energies where the suppression by large charge densities, leaves a relatively weakly interacting, $\Jj \sim e^{-\tilde{\Qq}^2}J $, liquid phase. The extreme of this is seen by considering a different rescaling 
	\begin{equation}
	T =\bar{T} q^{-2}, \quad \mu = \bar{\mu} q^{-2}, \label{scaling2}
	\end{equation} %$\Ss = \beta\mu \Qq - \ln\sqrt{1/4 - \Qq^2}$ $\ln\sqrt{1/4 - \Qq^2} = \mu \Qq -T\Ss$
	where the system transitions to a finite non-rescaled charge density $\Qq = 0\to \sqrt{1-e^{-4 J/\bar{T}}} /2$, shown in App. \ref{cl2}. 
	This suppresses the effective interaction $\Jj = e^{-qJ/\bar{T}} J\to 0$, yielding a free integrable ($v \to 0$) system. As such, small perturbations, stemming from $\bar{\mu}$, to the $\Qq=0$ symmetric-Majorana state, induces a jump to a Fermi gas at finite (positive or negative) charge density, hence breaking the U$(1)$ symmetry. This transition is thus from a maximally chaotic to a non-chaotic state.
	Such a Fermi gas has an entropy $\Ss = -\beta\mu \Qq - \ln\sqrt{1/4 - \Qq^2}$.
	To leading order in $\bar{T} \equiv \bar{\beta}^{-1}$, this indicates a drastic drop in entropy $\ln 2\to\bar{\beta} J e^{-4 \bar{\beta} J}$. Such an instability is also seen in RN black holes at low temperatures \cite{Hartnoll2008Dec,Karahasanovic2012Apr}. The RN transition is from a large black hole to a small one. Since the Bekenstein-Hawking entropy is proportional to the surface area, this also corresponds to a drop in entropy. Lastly, both RN and SYK transitions also include an unstable solution, with a negative bulk modulus.

	For $\bar{T} = o(q^0)$, there is a first-order quantum phase transition from $\Qq=0$, to maximum density $\Qq= 1/2$, at $\bar{\mu}_0 = 4J$. If $\bar{\mu} <\bar{\mu}_0$, then we are left with a Majorana SYK ground state solution with an extensive entropy. Such a finite entropy, at $T=0$, is also the defining property of RN metals \cite{Zaanen2015Nov}. If $\bar{\mu} > \bar{\mu}_0$, we are left with a zero entropy harmonic oscillator vacuum state. This first-order quantum phase transition is also observed in the finite $q$ equivalent models \cite{Azeyanagi2018Feb,Ferrari2019Jul}. Such a transition is again related to gravity, this time the classical HP transition \cite{Hawking1982Jan}, which has a large black hole to (non-interacting) thermal radiation, with zero entropy, transition. 
	
	Also of note, is the conjecture of black holes being the fastest scramblers \cite{Sekino2008Oct}, and as such chaotic \cite{Shenker2014Mar,Maldacena2016Nov,Engelsoy2016Jul}. Assuming this holds, the gravitational analogies extend over to a chaotic-to-chaotic RN transition, as we found in the scaling regime \eqref{scaling}. It would further include a chaotic-to-non-chaotic HP transition, where the non-chaotic phase is (non-interacting) thermal radiation,  corresponding to our observed low-temperature crossover \eqref{scaling2}. 
	
	It is quite remarkable that there are at least two RN models which also qualitatively match our phase diagram \cite{Chamblin1999Aug,Dolan2016May} by terminating at a first order phase transition at $(q_B,\beta) = (0,\beta_Z)$. This is reminiscent of how our coexistence line terminates at $(\bar{T},\bar{\mu}) = (0,\bar{\mu}_0)$. This is in contrast to extended space RN black holes and vdW, with coexistence lines extending to the point $(0,0)$. At the other end, both models terminate at a second order transition. Of note is that \cite{Chamblin1999Aug} has the same effective exponents matching II.\hyperref[a]{(a)} \cite{Niu2012Jan}. %The analogy $(T,\mu,\Qq) \leftrightarrow (q_B,\beta,r)$ between our model and the RN black hole in  also yields interesting results.
	
	All these similarities are perhaps not so surprising from the perspective of holography. This is because the SYK model is a $(0+1)$-dimensional conformally symmetric theory at low temperatures. As such, from the AdS/CFT correspondence, one would conjecture that it is a CFT on the boundary of some $\AdS_{1+1}$ space. Standard $(1+1)$ dimensional gravity is topological and displays only trivial physics, hence we consider non-standard gravity, the simplest of which are the Jackiw–Teitelboim (JT) black holes. They may be viewed as the dimensional reduction, or the near-horizon theory of near-extremal (minimal mass) higher-dimensional black holes \cite{Nayak2018Sep,Moitra2019Nov,Moitra2019Jul}. One such model \cite{Cao2021Mar} even has a phase transition with calculated effective exponents matching that of II.\hyperref[b]{(b)} and $\upbeta = 1/2$. %$(q_B^2,r) \leftrightarrow (\mu,\Qq)$
	
	%https://physics.stackexchange.com/questions/27525/do-thermodynamic-quantities-in-cft-correspond-to-something-different-in-ads-cft/27526#27526
	%\jl{Note that $\sgn(C_P) = \sgn(K)$}

	%Analogies between such models and the SYK model have been studied with some success \cite{Gaikwad2020Feb}.

	\emph{Conclusion.---} 
	We presented an analytic study of the complex large-$q$ SYK model, showing that it displays an RN-like phase transition.  Prior numerical analyses of the finite $q$ case have observed that the phase diagram scales away at larger values of $q$ \cite{Azeyanagi2018Feb,Ferrari2019Jul}. We showed that if one considers rescaled quantities as described by \eqref{scaling} and \eqref{scaling2}, then the transition in fact still exists at infinite $q$.
	
	One can further study the overlap of our large $q$ results with that of the finite $q$ numerical results \cite{Ferrari2019Jul}. A natural choice is to consider the relative error between their respective critical values. Such an analysis is provided in App. \ref{AppOverLap}, where we found relative errors which appear to converge to zero rather quickly as $q$ increases. This supports the relevance of the $q\to \infty$ limit for finite $q$ models.
	
	In contrast to the finite $q$ case, which has asymmetric (differing over the coexistence line) irrational exponents \cite{Azeyanagi2018Feb,Ferrari2019Jul}, we found symmetric rational numbers. One should note that Refs. \cite{Azeyanagi2018Feb,Ferrari2019Jul} actually determined effective exponents because the ordering field was assumed to be the chemical potential. However, as we have seen in our analysis, field mixing needs to be taken into account to determine the universality class and the critical exponents. The only exception where field mixing plays no role is $\upbeta$. The small deviation from the mean-field value $1/2$ in Refs. \cite{Azeyanagi2018Feb,Ferrari2019Jul} might be due to numerical error, and as such whether the finite $q$ SYK model also falls into the vdW universality class remains an open question.
	
	By comparing the critical exponents to other models, we found that the complex large-$q$ SYK model and many RN black holes find themselves in the same mean-field vdW universality class.

	Further, in the low reduced temperature regime, defined by the scaling \eqref{scaling2}, we found a jump between maximally chaotic and non-chaotic phases, also observed in generalized/coupled SYK models \cite{Luo2019Jun,Maldacena2018Apr,Garcia-Garcia2018Jun,Kim2021Mar,Klebanov2020Nov,Sahoo2020Oct}. The coexistence line dividing the two phases terminates at a first-order quantum phase transition from a Majorana ground state to a Fermi gas, hence a drop from non-zero residual entropy down to zero. This feature is shared with the first-order Hawking-Page (HP) transition between a large black hole and thermal radiation \cite{Hawking1982Jan}. As such, the gravitational analogies extend to the low-temperature regime. 
	
	From the perspective of AdS/CFT, these similarities between our model and charged black holes are perhaps not too surprising. This is because the SYK model is conformally symmetric in the infrared limit \cite{Maldacena2016Nov}. However, the details narrow down the list of possible gravity duals to the SYK model \cite{Rosenhaus2019Jul}.  The analytical expressions derived in this work, the power laws, equation of state, and grand potential, serve as a guide towards finding this dual. 
	
	We conclude with the natural question of whether any columns in table \ref{tab:title} or other mentioned analogies are part of an AdS/CFT dictionary. In other words, is there (asymptotic) equivalence between any of the partition functions?\\
	
	We would like to thank Peter Sollich for helpful discussions on field mixing, critical exponents and scaling relations. This work was funded by the Deutsche Forschungsgemeinschaft (DFG, German Research Foundation) - 217133147/SFB 1073, project~B03 and the Deutsche akademische Austauschdienst (DAAD, German Academic Exchange Service).
	\bibliography{ref2}

	\appendix
	
	\section*{Appendix}

	%Supplementary material \\[1ex] \normalsize additional mathematical detail to ``The shared universality of charged black holes and the many many-body SYK model''
	\setcounter{table}{0}
	\renewcommand{\thetable}{A\arabic{table}}%
	\renewcommand{\thesection}{A.\arabic{section}}
	\renewcommand{\theequation}{A\arabic{equation}}
	\setcounter{figure}{0}
	\setcounter{equation}{0}
	\renewcommand{\thefigure}{A\arabic{figure}}
	
	\section{Validity of q re-scaling \label{cl1}}
	
	The thermal Green's function
	$\Ggg(\tau-\tau')  \equiv \Ggg(\tau,\tau') = -\Tt \ex{c(\tau) c^\dag(\tau')}$ is the solution to Dyson's equation
	\begin{align}
		[\Ggg -\Ggg_0](\tau,\tau')= & \nint[0][\beta]{\tau_1} {\small \text{$\mathrm{d}t$}} \Ggg(\tau,t) \Sigma(t,\tau_1) \Ggg_0(\tau_1,\tau'),\label{ImDy}
		%\\[\Ggg -\Ggg_0](\tau',\tau)=&\nint[0][\beta]{\tau_1} \nint[0][\beta]{\tau_2} \Ggg_0(\tau',\tau_2) \Sigma(\tau_2,\tau_1) \Ggg(\tau_1,\tau),
	\end{align} %$c(\tau) \equiv e^{\Hh \tau} c e^{-\Hh \tau}$
	with non-interacting Green's function $\Ggg_0(\tau) = \ex{c^\dag c}_0 - \Theta(\tau)$ and self energy $\Sigma$. For the $q/2$-body interacting SYK model $\Sigma(t) = \Ggg(t) 2 J^2 [-4\Ggg(t)\Ggg(-t)]^{q/2-1}/q$ \cite{Fu2018}. 
	We write $\Ggg(\tau) = [\Qq - \sgn(\tau)/2] e^{\Delta \ggg(\tau)}$, with $\Delta \equiv 1/q$, leaving $q\Sigma = 2\Jj^2 e^{(1-2\Delta)\ggg_+}\Ggg$. Here we have split $\ggg$ into symmetric/asymmetric parts $\ggg_\pm(-\tau) = \pm \ggg_\pm(\tau)$ and defined the charge density $\Qq \equiv \frac{1}{N}\sum_{i}\ex{c_i^\dag c_i} - 1/2$.  \\
	\begin{claim}{1} %, where $c_\tau = e^{\tau [\Hh,\cdot]} c$,
		In the large $q$ limit, we \emph{claim} that, for any thermodynamic variables which scale sub-exponentially in $q$ the solution to \eqref{ImDy} is given by $\ggg_-(\tau) = 2\Qq \dot{\ggg}_+(0)\tau$ and
		\begin{equation}
			e^{[1-2\Delta]\ggg_+(\tau)} = \frac{[\pi v/\lambda]^2}{\cos^2[\pi v(1/2-|\tau|/\beta)]}, \quad \frac{\pi v/\lambda}{\cos[\pi v/2]} = 1, \label{closure}
		\end{equation}
		where $\lambda = \beta \Jj \sqrt{1-2\Delta}$ and $\Jj \equiv [1-4\Qq^2]^{(q-2)/4} J$. Note that for non-rescaled variables, they take on the standard known forms given for neutral charge $\Qq=0$ in \cite{Maldacena2016Nov} or at finite charge in \cite{Louw2022Feb}.
	\end{claim}
	
	\emph{Proof:} 
	We take the approach of substituting the claimed solutions $\ggg_{\pm}$ into the full Dyson equation. We then gather the non-zero (error) terms $\Delta R_i$ and show that, given sub-exponentially scaling, $\Delta R_i \xrightarrow{\Delta\to0} 0$.  Using \eqref{ImDy}, $q\p_{\tau'}\ln\Ggg(\tau,\tau')\mid_{\tau'=0}$, for $\tau \ge 0$, reduces to
	\begin{equation}
		\dot{\ggg}(\tau) - \Jj^2\nint[0][\beta]{t} [\sgn(\tau-t)-2\Qq ] e^{\ggg_+(t) + \Delta \varphi_{\tau}(t)} = 0\label{gdot}
		%\dot{\ggg}(\pm \tau) = \Jj^2\nint[0][\beta]{t} [\pm \sgn(\tau-t)-2\Qq ] e^{\ggg_+(t) + \Delta \varphi_{\pm \tau}(\pm t)} \label{gdot}
	\end{equation} % \dot{\Ggg}_0(\tau) = -\delta(\tau)$
	with $\varphi_{\tau}(t) \equiv \ggg(\tau-t) - \ggg(\tau) - \ggg(-t)$. By differentiating again we obtain the two equations 
	%\begin{equation}
	$$\ddot{\ggg}_+ = 2\Jj^2 e^{[1-2\Delta]\ggg_+} [1+ \Delta R],\quad \ddot{\ggg}_-= \Delta 2\Qq \Jj^2 R_3,$$ %\label{LiouLead}
	%\end{equation} 
	with $R \equiv R_1 + R_2$. Ignoring the error terms, we have $\ggg_-(\tau) \sim \epsilon \tau$, while $\ddot{g}_+$ reduces to a Liouville equation with solution \eqref{closure}.  The boundary condition in \eqref{closure} enforces $\Gg(0^+) =\Qq - 1/2$. Substituting the solutions back into \eqref{gdot} one finds error terms
	\begin{align}
		R_1(\tau) &= - [ e^{-(1/2-\Delta)\ggg_+(\tau)}\dot{\ggg}_+(\tau)/\Jj]^2/2 \label{R1}\\
		R_2(\tau) &= e^{-(1+\Delta)\ggg_+(\tau)}  \frac{\Ii_{\tau}(t)\vert^\tau_{0} - \Ii_{\tau+\beta}(t)\vert^\beta_{\tau}}{2} \label{R2}\\
		R_3(\tau) &= - [\, \Ii_{\tau}(t)\vert^{\tau}_{0} + \Ii_{\tau+\beta}(t)\vert^{\beta}_{\tau}] -\frac{\dot{\ggg}_+(\tau)}{\Jj}\frac{\dot{\ggg}_-(\tau)}{2\Qq \Jj} \label{R3}
	\end{align}
	where we have defined the indefinite integral
	\begin{equation}
		\Ii_\tau (t) \equiv\nint{t}  \dot{\ggg}_+(\tau-t) e^{\Delta \ggg_+(\tau-t) +(1-\Delta) \ggg_+(t)}. \label{indef}
	\end{equation}
	%Here we have used the Kubo-Martin-Schwinger  (KMS) relation $\ggg_+(-|t|) = \ggg_+(\beta-|t|)$ to rewrite the integrands.
	
	We would next like to find the $q$-dependent scaling conditions on $\lambda$ for which all $\Delta R_i \xrightarrow{\Delta\to0} 0$. Using \eqref{closure}, the bound $|R_1|\lesssim 2$ follows from
	$$\frac{\dot{\ggg}_+(\tau)}{\Jj}  = - \frac{2\sin(\pi v(1/2-\tau/\beta))}{\sqrt{1-2\Delta}} e^{(1/2-\Delta)\ggg_+(\tau)}.$$ %- \frac{\pi v \tan(\pi v(1/2-\tau/\beta))}{\lambda \sqrt{1-2\Delta}}
	
\begin{figure}
		\centering
		\includegraphics[width=0.7\linewidth]{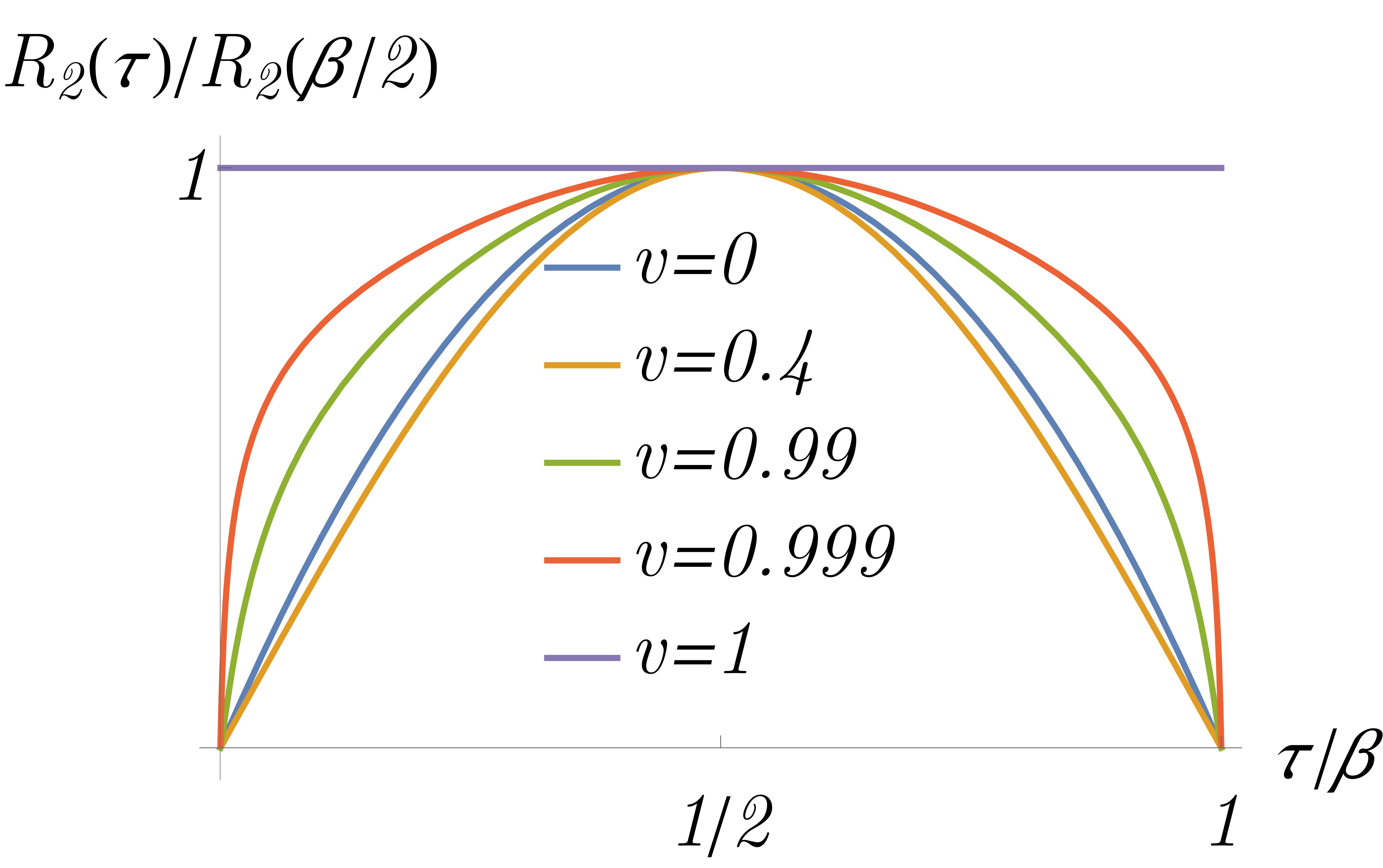}
		\caption{The error plotted for various values of $v$. \hspace*{\fill}}
		\label{fig:rorm}
	\end{figure}

	To bound \eqref{R2} and \eqref{R3} we first evaluate the integral \eqref{indef}. To simplify the analysis we would like to replace exponentials like $e^{(1+\Delta) \ggg_+}$ with $e^{\ggg_+}$, under the integral, which is justified if the corresponding function, e.g., $e^{\Delta \ggg_+}$, remains differentiable under said limit. To see when this holds, we note that $(\pi v/\lambda)^{2 \Delta} \le e^{\Delta (1-2\Delta)\ggg_+(t)} \le 1 $ saturates for $\vert\ln [v/\lambda] \vert\lesssim q$. In this case, \eqref{indef} evaluates to
	\begin{align}
		\Ii_\tau(t) %&=-\lambda \nint{t}  \frac{2\sin[\pi v(1/2-(\tau-t)/\beta)]}{\beta (1-2\Delta)} e^{\ggg_+(\tau-t)/2 +\ggg_+(t)}\\
		=&  \nint{y}  \tan[\pi v/2-(x-y)] \frac{-2\cos^2(\pi v/2)}{\cos^2[\pi v/2-y]} \notag\\
		=& 2\cot[\pi v-x] \sin y \frac{\cos[\pi v/2]}{\cos[\pi v/2-y]} \notag\\
		&+ 2\left[\frac{\cos(\pi v/2)}{\sin(\pi v-x)}\right]^2 \ln \frac{\cos[\pi v/2-(x-y)]}{\cos(\pi v/2-y)}, \label{Ii}
	\end{align}
	where $x =  \pi v\tau/\beta$ and $y = \pi v t/\beta$. This yields the error \eqref{R2} 
	\begin{align*}
		R_2(\tau) %&= e^{-\ggg(\tau)}\frac{f(\beta-\tau) +f(\tau)}{2}\\
		%\sim& \frac{\cot(x) \sin(\pi v -x)+\cot(\pi v -x) \sin(x)}{2 e^{\ggg_+(\tau)/2}}\\
		%&+ \left[\frac{\ggg_+(\tau)}{\sin^2(x)} + \frac{\ggg_+(\tau)}{\sin^2(\pi v -x)}\right]\frac{e^{\ggg_+(\beta/2)-\ggg_+(\tau)} }{2}\\
		\sim &  \bigg[\frac{\csc^2[\pi v-x] + \csc^2 x}{2} \ggg_+(\tau) - 1 \notag\\
		&\quad+ \frac{2\sin^2[\pi v/2]}{\sin[\pi v-x]\sin x}\bigg] 2\cos^2[\pi v/2-x]. 
		%\sim &  -\frac{2\csc(x-\pi v)\csc(x)}{2 } \frac{\ggg(\tau)}{\sec^2(\pi v/2-x)} - \cos^2(\pi v/2-x) \\
		%& + \frac{[\csc(x-\pi v) + \csc x]^2}{2} \left[\sin(\pi v-x) \sin x+\frac{\ggg(\tau)}{\sec^2(\pi v/2-x)}\right]\\
	\end{align*}
	%R(\tau)\sim &  \frac{[\csc^2[\pi v-x] + \csc^2 x]\ggg(\tau) +4\sin^2[\pi v/2]\csc[\pi v-x]\csc x}{\cos^2[\pi v/2]}  e^{-\ggg_+(\tau)}-2. 
	which, seen in fig. \ref{fig:rorm}, has a maximum at $\tau =\beta/2$ given by
	$$R_2(\beta/2) = 2 + 2  \frac{\ln[\pi v/\lambda]^2}{1-[\pi v/\lambda]^2}.$$
	
	The final error term \eqref{R3} is bounded by using \eqref{gdot} to write
	$$\dot{\ggg}_-(\tau) \xrightarrow{\Delta\to 0} -2\Qq \Jj^2\nint[0][\beta]{t} e^{\ggg(t)} = -\Qq \nint[0][\beta]{t} \ddot{\ggg}_+(t),$$
	which integrates to $\dot{\ggg}_-(\tau)  = 2\Qq\dot{\ggg}_+(0)$, hence matching the postulated solution. Together with \eqref{Ii}, one may show that this leaves an error with maximum magnitude of $4 \sin^2(\pi v/2)$ at $\tau = 0$. 
	
	With all three of these error terms, one notices that $\Delta R_i \xrightarrow{q\to \infty} 0$ as long as large $\vert\ln [v/\lambda] \vert\lesssim q$. For large $\lambda$, the relation \eqref{closure} implies that $v \sim 1$. This means our solutions remains valid for any $\ln\beta \Jj \ll q$, which includes the polynomial scalings $\beta\Jj \sim q^{\alpha}$ considered in the work, i.e, $T = \tilde{T} q^{-1}$ and $T =\bar{T} q^{-2}$, with $J = \Oo(q^0)$.\\
	
	\section{Low temperature charge transition \label{cl2}}
	
	\begin{claim}{2} 
		At small reduced temperature $T_r \equiv \tilde{T}/\tilde{T}_\text{crit}$, where $\tilde{T}_\text{crit} = 2 e^{-3/2} J$, the charge density jumps as
		\begin{equation}
			\Qq_g = \frac{\cosh^{-1} [e^{2\bar{\beta} J}] }{q 2 \bar{\beta} J} \to \Qq_\ell= \frac{\sqrt{1-e^{-4 \bar{\beta} J}}}{2}. \label{barQg} % \sim 1+ \frac{\bar{T} \ln 2}{2J}
		\end{equation}
		For convenience, we have defined $\tilde{T} \equiv q^{-1} \bar{T}$, $\tilde{\mu} \equiv q^{-1/2} \bar{\mu}$.
		
		As a special case, if we consider $\bar{T},\bar{\mu}$ fixed in the large $q$ limit, we find a transition from $\Qq = 0 \to \sqrt{1-e^{-4 \bar{\beta} J}}/2$, i.e., a transition to a finite non-rescaled charge density as discussed in the work.
	\end{claim}

	\emph{Proof:} Consider the  grand and chemical potentials %$\tilde{\Qq}_c \sim \sqrt{3/2 \jl{-} \ln T_r + \ln(1 - \ln T_r)}$
	\begin{equation}
		\tilde{\Omega} = -2\Jj-2 \tilde{\Qq}^2 [2\Jj+\tilde{T}], \quad \tilde{\mu} = 4 \tilde{\Qq} [\Jj+\tilde{T} ],\label{OGGrandPot}
	\end{equation}
	at low reduced temperature. In other words, consider the asymptotic behavior for $\tilde{T} \lll J$ of the charge densities on both sides of the coexistence line. Though we do not have explicit forms for these charge densities, we can gain some insight by considering fig.\!\!\jl{1} from the work. Here we note that gaseous solution must be smaller than the charge density at the first turning point of $\tilde{\mu}(\tilde{\Qq})$, $\tilde{\Qq}_g < \tilde{\Qq}_b$, while the liquid charge density must be larger than the second turning point $\tilde{\Qq}_\ell >\tilde{\Qq}_c $. These turning points are
	\begin{equation}
		\tilde{\Qq}_b = \sqrt{\frac{1}{2} - \www_0\left[\frac{- T_r}{e}\right]},\, \tilde{\Qq}_c = \sqrt{\frac{1}{2} - \www_{-1}\left[\frac{- T_r}{e}\right]},\label{tps}
	\end{equation}
	where $\www(x)$ is the product log satisfying $\www e^{\www} = x$. Here the subscripts indicate the various branches, with $\www_0$ corresponding to the principal branch.

	On the gaseous side we have $\tilde{\Qq}_b = \sqrt{1/2} +\Oo(T_r)$, implying relatively small charge densities $\tilde{\Qq}_g < \sqrt{1/2}$. As such, the strong coupling $\Jj_g \equiv J e^{-\tilde{\Qq}_g^2} = \Oo(T_r^0)$ dominates in \eqref{OGGrandPot},
	\begin{align} 
		\tilde{\Omega}_g &=-2 \Jj_g-4\tilde{\Qq}_g^2[ \Jj_g+\Oo(T_r)], & \tilde{\mu}_g &= 4 \tilde{\Qq}_g [\Jj_g +\Oo(T_r)].    \label{strongPot}
	\end{align}
	
	For the liquid phase we have $\tilde{\Qq}_c = \sqrt{\ln[\beta_r \ln\beta_r ]  } +\Oo(1),$
	with $\beta_r \equiv 1/T_r$. As such, we have a relatively large charge density $\tilde{\Qq}_\ell > \tilde{\Qq}_c $ hence a large suppression in the coupling $\Jj_\ell \lesssim \tilde{T}/\ln \beta_r$. With this  \eqref{OGGrandPot} reduces to
	\begin{align}
		\tilde{\Omega}_\ell &= - 2 \tilde{T} \tilde{\Qq}_\ell^2[1 + \Oo(\Jj_\ell/\tilde{T})], & \tilde{\mu}_{\ell} &= 4 \tilde{\Qq}_{\ell} \tilde{T} [1 + \Oo(\Jj_\ell/\tilde{T})]. \label{weakPot}
	\end{align}
	where $\Jj_\ell/\tilde{T} = \Oo(1/\ln \beta_r)$.
	Here, the weakly interacting phase \eqref{weakPot} is in fact the leading order solution to free fermions
	\begin{equation}
		\tilde{\Omega}_0 = \frac{\bar{T}}{2} \ln[1 - 4 \Qq_\ell^2] ,  \quad\tilde{\mu}_0 = 2 \frac{\bar{T}}{\sqrt{q}} \tanh^{-1}[2 \Qq_\ell], \label{fullweakPot}
	\end{equation}
	i.e., expanding \eqref{fullweakPot} for small non-rescaled charge densities $\Qq \equiv \tilde{\Qq}/\sqrt{q}$, yields \eqref{weakPot} to leading order.
	These are the full solutions at large charge densities, since this yields small effective interactions.  An analysis using \eqref{weakPot}, while valid at infinite $q$, for fixed tilde'd variables, yields the incorrect zero temperature limit expressions for large finite $q$. As such, we focus on phase transitions from \eqref{strongPot} to \eqref{fullweakPot}, which includes the previous analysis as a solution under the appropriate limit. The phase transition occurs at the point of equal grand and chemical potential. Equating the expressions in \eqref{strongPot} and \eqref{fullweakPot}, yields the equations
	$$\ln(1-4\Qq_\ell^2) = - 4 \bar{\beta} \Jj_g [1+2\tilde{\Qq}_q^2], \quad 2 \Qq_\ell = \tanh[2 \bar{\Qq}_g \bar{\beta} \Jj_g],$$
	where $\tilde{\Qq} \equiv q^{-1/2}\bar{\Qq}$. The solution to these two are the roots of
	\begin{align}
		F(\tilde{\Qq}_g) &=  1+2 \tilde{\Qq}_g^2 - \frac{\ln \cosh\left[2 \bar{\beta}  \Jj_g\bar{\Qq}_g\right]}{2\bar{\beta}  \Jj_g}. \label{froot}%\\
		%&= 1+2 \tilde{\Qq}_g^2-\bar{\Qq}_g-\frac{1}{2\bar{\beta} \Jj_g}\ln \frac{1 + e^{-4 \bar{\beta} \Jj_g \bar{\Qq}_g}}{2}
	\end{align}
	As $\bar{T} \to 0$, the root is at $\bar{\Qq}_g = 1+2 \tilde{\Qq}_g^2 \sim 1$. Together with \eqref{strongPot}, this suggests that there is a first-order transition at $\bar{\mu}_0 = 4 J$. Above zero temperature, for any $\tilde{\Qq}_g = o(q^0)$, we simply have  $\Jj_g \sim J$ and as such the root of $F$, $\tilde{\Qq}_g$, and the corresponding liquid charge density are as given in  \eqref{barQg}. \\
	
		\section{Scale invariant grand potential \label{cl3}}
	
	\begin{claim}{3} 
		Consider the shifted rescaled grand potential from the work
		\begin{equation}
			f  = -\rho^2 [3\rho^2/4 +t/3]+\Oo(\rho^5). \label{ff}
		\end{equation}
		Here $\rho$ is the solution to the equation of state \eqref{weakPot} around the critical point
		\begin{equation}
			m-2t/3 = \rho (2t/3 +\rho^2) +\Oo(\rho^4)\label{EOSSM}
		\end{equation}
		with $m,t$ and $\rho$ defined in the work.
		We claim that in terms of the ordering field $\hh = m-2 t/3$ that \eqref{ff} is homogeneous
		$f(t,\hh) = t^{2} f(1,\hh |t|^{-3/2})$. 	
	\end{claim}

	\emph{Proof:} The above function $f$ satisfies $\p_m f = -\rho$. We may rewrite $f$ in terms of $\rho \equiv - |t|^{1/2} \dot{\Psi}$, as
	$f = |t|^2\Psi$, where for $t<0$
	$$\Psi \equiv -\dot{\Psi}^2 [3\dot{\Psi}^2/4-1/3]+\Oo(\rho^5).$$
	To prove homogeneity, we must show that $\dot{\Psi}$ is a function only of $\omega \equiv \hh |t|^{-3/2}$, hence leaving $\Psi(\omega) \equiv f(1,\omega)$.
	From \eqref{EOSSM}, we have the relation $\omega = \dot{\Psi} (\dot{\Psi}^2 - 2/3)$. This implies that $\dot{\Psi}$ is indeed purely a function of $\omega$. One may further show that $\p_{\omega}\Psi(\omega) = \dot{\Psi}(\omega)$. The liquid phase solution is 
	\begin{align*}
		\dot{\Psi}_\ell(\omega)
		%&= \frac{1}{\sqrt{3} (y+\sqrt{y^2-1})^{1/3}} + \frac{(y+\sqrt{y^2-1})^{1/3}}{\sqrt{3} }
		&= \begin{cases}
			[2/3]^{1/2}+  3 \omega/4 - (3/2)^{7/2} \omega^{2}/4 + \Oo(\omega^3)\\
			\omega^{1/3} + 2\omega^{-1/3}/9 + \Oo(\omega^{-5/3}).
		\end{cases}
	\end{align*}
	
	Over the range $\dot{\Psi} \le 2^{3/2}/3$, $\omega(\dot{\Psi})$ is three-to-one, with the two remaining solutions: $\dot{\Psi}_{\text{b}-\text{c}}(w) = \dot{\Psi}_\ell(-\omega) - \dot{\Psi}_\ell(\omega)$ and $\dot{\Psi}_{g}(\omega) = -\dot{\Psi}_\ell(-\omega)$, where the latter corresponds to the gaseous phase. Explicitly
	$$ 	\Psi_\ell(\omega) =  
	\begin{cases}
		%-(2 + |2/3|^{-3/2} \omega|)^2/9+ 1/3 + \Oo(\omega^3)\\
		-1/9 - \sqrt{2/3} \omega- 3 \omega^2/8 + \Oo(\omega^3)\\
		%-(1 +3 (|2/3|^{-3/2} \omega)^{2/3})^2/27 - \frac{1}{27} +\Oo(\omega^{-2/3})
		-\frac{3}{4} \omega^{4/3} - \frac{\omega^{2/3}}{3} - \frac{2}{27} + \Oo(\omega^{-2/3}).
	\end{cases}$$

	Considering the grand potential, one notes that the gaseous solution is thermodynamically preferred for $\omega<0$ while the liquid solution is preferred for $\omega>0$, i.e., $\dot{\Psi}^*(\omega) = \sgn(\omega) \dot{\Psi}_\ell(|\omega|)$.
	
	\section*{Overlap with finite \lowercase{q} model \label{AppOverLap}} 
	
\begin{figure}[h]
		\centering
		\includegraphics[width=0.9\linewidth]{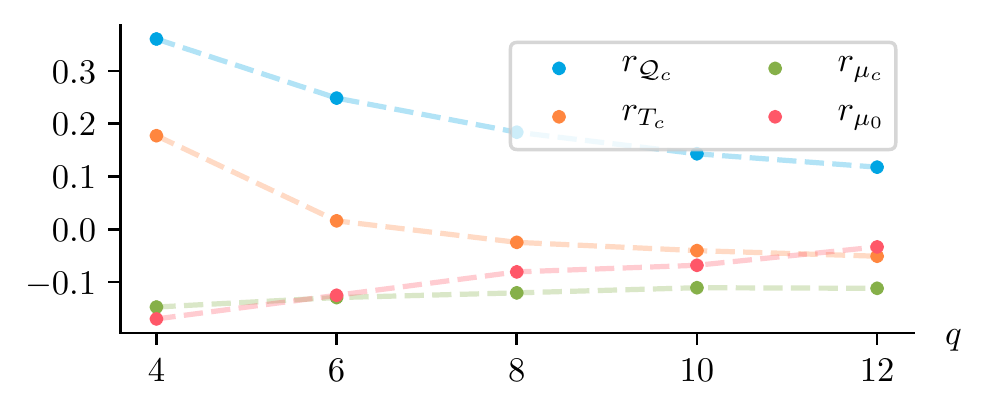}
		\caption{Plots of the relative errors \eqref{relEr} between the large $q$ estimated critical points and the numerically calculated critical points for various finite $q$ values.\hspace*{\fill}}
		\label{fig:relerrorscrit}
	\end{figure}

	We can test the extent to which our large $q$ results capture the finite $q$ physics, by comparing our derived quantities to the numerically derived ones in \cite[Table I.]{Ferrari2019Jul}. We focus on the relative error 
	\begin{equation}
		r_{x} = \frac{x}{x^{(q)}} - 1 \label{relEr}
	\end{equation}
	where $x$ is the large $q$ estimation, while $x^{(q)}$ corresponds to the numerical results. The particular quantities we compare are the critical values
	\begin{align*}
		\Qq_c =  \sqrt{\frac{3}{2 q}}, \quad T_c = \frac{2 J e^{-3/2}}{q} \quad \mu_c = 6\Qq_c T_c, \quad \mu_0 = \frac{4 J}{q^2}
	\end{align*} 
	derived in the work. We choose a matching coupling constant to \cite{Ferrari2019Jul}, $J =  \sqrt{q 2^{1-q}}$. As such, all the critical values become functions of $q$, which we compare with their numerically derived values in Fig. \ref{fig:relerrorscrit}. 
	
	The comparison is given for $q \in \{4,6,8,10,12\}$. From this, we note that the estimated values appear to converge to their finite $q$ values rather fast as $q$ increase. In order to reduce the relative error at smaller values of $q$, for instance $q=4$, one would have to consider $1/q^2$ corrections to the Green's functions \cite{Tarnopolsky2019}.

\end{document}